\documentclass[runningheads]{llncs}
\usepackage{graphicx}
\usepackage{booktabs}
\usepackage{amsmath}
\usepackage{amssymb}
\begin{document}
\title{A$^3$DSegNet: Anatomy-aware artifact disentanglement and segmentation network for unpaired segmentation, artifact reduction, and modality translation}
\titlerunning{Anatomy-aware artifact disentanglement network}
%

\author{Anonymous}
	\author{Yuanyuan Lyu\inst{1} \and
	Haofu Liao\inst{2} \and
	Heqin Zhu\inst{3,5} \and 
	S. Kevin Zhou\inst{3,4,5}}
	\institute{Z$^2$Sky Technologies Inc., Suzhou, China \and 
	Department of Computer Science, University of Rochester, NY, USA \and 
	Medical Imaging, Robotics, and Analytic Computing Laboratory and Engineering (MIRACLE) Group \and 
    School of Biomedical Engineering \& Suzhou Institute for Advance Research, University of Science and Technology of China, Suzhou, 215123, China \and 
    Key Lab of Intelligent Information Processing of Chinese Academy of Sciences (CAS),
    Institute of Computing Technology, CAS, Beijing, 100190, China \email{s.kevin.zhou@gmail.com} }

\maketitle              
\begin{abstract}
Spinal surgery planning necessitates automatic segmentation of vertebrae in cone-beam computed tomography (CBCT), an intraoperative imaging modality that is widely used in intervention. However, CBCT images are of low-quality and artifact-laden due to noise, poor tissue contrast, and the presence of metallic objects, causing vertebra segmentation, even manually, a demanding task. In contrast, there exists a wealth of artifact-free, high quality CT images with vertebra annotations. This motivates us to build a CBCT vertebra segmentation model using unpaired CT images with annotations. To overcome the \textit{domain and artifact gaps} between CBCT and CT, it is a must to address the \textit{three heterogeneous tasks} of vertebra segmentation, artifact reduction and modality translation all together. To this, we propose a novel \textit{anatomy-aware artifact disentanglement and segmentation network}  (\textbf{A$^3$DSegNet}) that intensively leverages knowledge sharing of these three tasks to promote learning. Specifically, it takes a random pair of CBCT and CT images as the input and manipulates the synthesis and segmentation via different decoding combinations from the disentangled latent layers. Then, by proposing various forms of consistency among the synthesized images and among segmented vertebrae, the learning is achieved without paired (i.e., anatomically identical) data. Finally, we stack 2D slices together and build 3D networks on top to obtain final 3D segmentation result. Extensive experiments on a large number of clinical CBCT (21,364) and CT (17,089) images show that the proposed \textbf{A$^3$DSegNet} performs significantly better than state-of-the-art 
competing methods trained independently for each task and, remarkably, it achieves an average Dice coefficient of 0.926 for unpaired 3D CBCT vertebra segmentation.

\keywords{Modality translation \and Unpaired segmentation \and Metal artifact reduction \and Disentanglement learning.}
\end{abstract}
\section{Introduction}

Cone-beam computed tomography (CBCT) has been widely used in spinal surgery as an intraoperative imaging modality to guide the intervention~\cite{zhou2021review,zhou2019handbook}. However, compared with conventional CT, intraoperative CBCT images have pronounced noise and poor tissue contrast \cite{schafer2011mobile,siewerdsen2011cone}. Moreover, it is common to have metallic objects (such as pedicle screws) present during operation, which cause metal artifacts and degrade the quality of CBCT images \cite{pauwels2013quantification}. 
To facilitate spinal surgery planning and guidance, it is of great importance to accurately identify the vertebrae~\cite{burstrom2019machine}; yet the poor CBCT image quality makes it challenging to delineate the vertebra shape even manually.

\begin{figure}[t]
	\centering
	\includegraphics[width=0.8\linewidth]{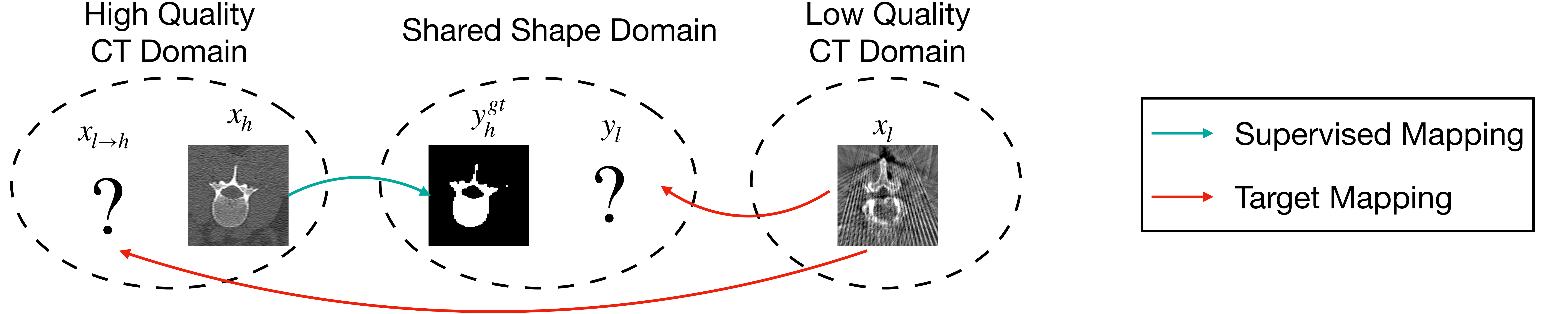}
	\caption{Sample images $x_l$ from the CBCT domain, $x_h$ from the CT domain and the paired vertebra segmentation $y_l^{gt}$. Due to the lack of high-quality, artifact-free CBCT images with vertebra annotations, it is challenging to directly learn a reliable CBCT vertebra segmentation model. We propose to leverage the knowledge from CT (both image and shape) to address this challenge under an unpaired setting.}
	\label{fig:pipeline}
\end{figure}

This paper aims to design a computational method to automatically segment vertebrae from clinical CBCT images, not from cadaver images as in \cite{pauwels2013quantification}. Since it is challenging to create a large number of CBCT images with annotations and yet high quality (artifact-free and high contrast) spinal CT datasets with vertebra delineations are easy to access \cite{yao2016multi}, we investigate the feasibility of learning a CBCT vertebra segmentation model using unpaired CT images with annotations as in Fig.~\ref{fig:pipeline}. Such learning has to overcome two obvious gaps: (i) the \textbf{modality gap} between CT and CBCT, that is, the image appearances look different even for the same content; and (ii) the \textbf{artifact gap} as the CT image is artifact-free and the CBCT is artifact-laden. In other words, we have to address \textit{three heterogeneous tasks of vertebra segmentation, artifact reduction, and modality translation} all together in order to derive a good solution.


There are existing methods that deal with modality translation ~\cite{huang2018multimodal,lee2018diverse,liu2017unsupervised,zhu2017unpaired}, or modality translation and artifact reduction~\cite{liao2019adn}, or modality translation and unpaired segmentation~\cite{kamnitsas2017unsupervised,Tsai_adaptseg_2018,zhang2018translating}. However, none of them can tackle all three. In this paper, we propose \textbf{for the first time} a unified framework that jointly addresses the three tasks, building on top of artifact disentanglement network (ADN)~\cite{liao2019adn}. Specifically, we propose a novel \textit{anatomy-aware artifact disentanglement and segmentation network} (\textbf{A$^3$DSegNet}) that 1) supports different forms of image synthesis and vertebra segmentation with joint learning, 2) utilizes an \textit{anatomy-aware de-normalization (AADE) layer} to boost the image translation performance by explicitly fusing anatomical information into the generator, and 3) induces different forms of consistency between the inputs and outputs to guide learning. Given unpaired CBCT and CT images, the proposed framework encodes disentangled representations and manipulates the synthesis and segmentation via different combinations of the decodings. Then, by discovering various forms of consistency among the synthesized images and among segmented vertebrae, self-learning from images and CT annotations is achieved without paired data for CBCT. Furthermore, to increase the segmentation performance, we utilize the anatomy-aware image translation to guide unpaired 3D segmentation for better inter-slice continuity by inducing more 3D shape consistencies. 

In summary, the contributions of this work are as follows:
\begin{itemize}
	\item By utilizing disentangled representations and anatomical knowledge from the target domain, we introduce a unified framework for tackling unpaired vertebra segmentation, artifact reduction, and modality translation, building on top of ADN. The three tasks benefit from each other via joint learning.
	\item We propose a novel A$^3$DSegNet that supports different forms of image synthesis and vertebra segmentation and discovers different forms of consistency to enable disentanglement learning. Also, we utilize an AADE layer to explicitly fuse the anatomical information, learned through shape consistency, into the generator and therefore boost the image synthesis performance.
	\item We embed the anatomy-aware disentanglement network as image translator into a final 3D segmentation network, which retains the spatial continuity between slices. 
	Ultimately, it achieves an average Dice coefficient of 0.926 for unsupervised 3D vertebra segmentation from CBCT.
\end{itemize}

\begin{figure*}[t]
	\begin{center}
		\includegraphics[width=0.9\linewidth]{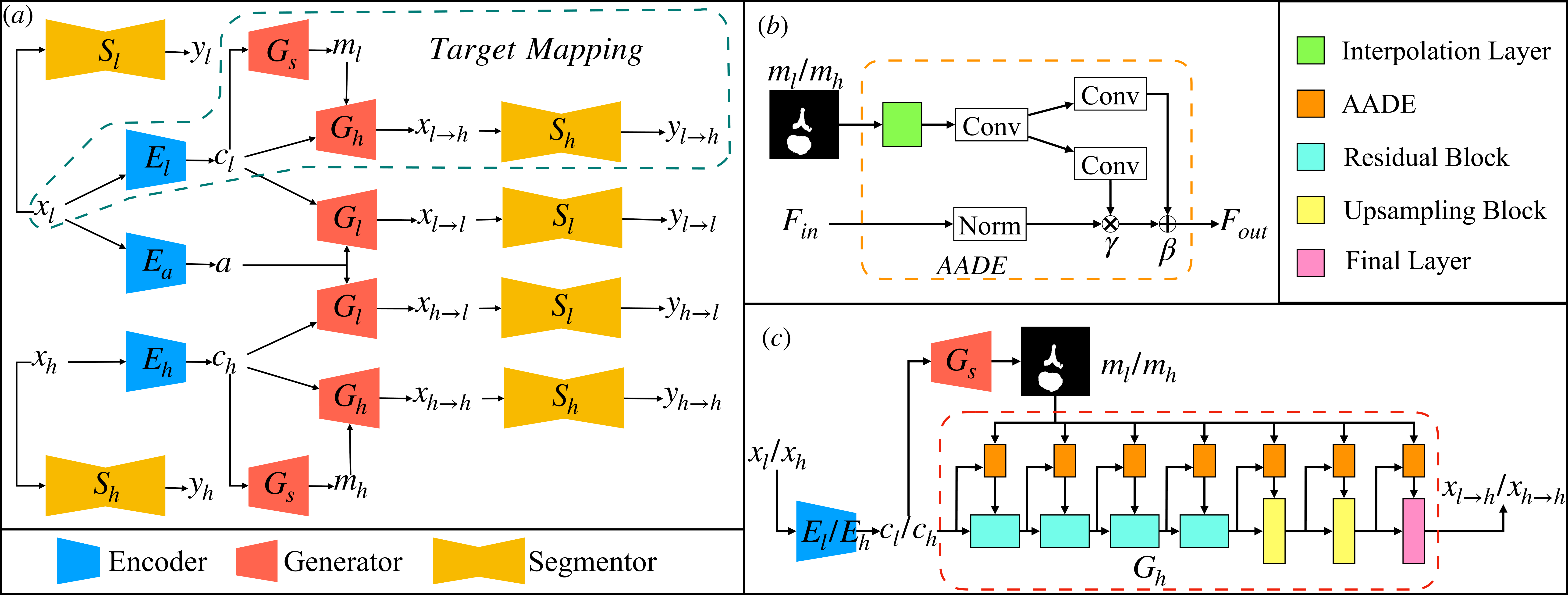}
	\end{center}
	\caption{(a) The architecture of anatomy-aware artifact disentanglement \& segmentation network (A$^3$DSegNet). (b) The proposed anatomy-aware de-normalization (AADE) layer. (c) The detail of anatomy-aware generator $G_h$.}
	\label{fig:network}
\end{figure*}

\section{Methodology}
Let $\mathbb{X}_l$ be the domain of low-quality, artifact-laden CBCT images, $\mathbb{X}_h$ be the domain of high-quality, artifact-free CT images, and $\mathbb{Y}$ be the domain of vertebra shapes. A CBCT image $x_l \in \mathbb{X}_l$ is usually noisy and may contain streak artifacts while a CT image $x_h \in \mathbb{X}_h$ is artifact-free and provides more anatomical details. A vertebra shape $y \in \mathbb{Y}$ can be presented as a binary segmentation mask where $y(k) \in \{0, 1\}$ indicates whether a pixel at location $k$ belongs to a vertebra. 

The proposed approach aims to learn a translator $\mathcal{F}: \mathbb{X}_l \rightarrow \mathbb{X}_h \times \mathbb{Y}$ that maps $x_l$ to its corresponding high-quality image $x_h \in \mathbb{X}_h$ and vertebra shape $y_l \in \mathbb{Y}$ without paired, anatomically identical groundtruth data $x_l$ and $y_l^{gt}$ available for supervision. To facilitate this unpaired learning, we assume the availability of a high-quality image dataset of $\{(x_h, y_h^{gt}) \mid x_h \in \mathbb{X}_h, y_h^{gt} \in \mathbb{Y}\}$. Fig.~\ref{fig:pipeline} shows sample images from these three domains. Note that $\mathbb{X}_l$ and $\mathbb{X}_h$ are independent, \textit{i.e.}, they are collected from different patients.


\subsection{Network architecture}\label{sec:architecture}

An overview of the proposed network architecture is shown in Fig.~\ref{fig:network}(a). Inspired by recent progress in disentangled image-to-image translation \cite{huang2018multimodal,lee2018diverse,liao2019adn}, we assume that the content $c_l$ (\textit{i.e.}, bones, soft tissues, etc.) and artifact $a$ (\textit{i.e.}, noises, streaks, etc.) of a low-quality image $x_l$ is disentangled in the latent space, see Fig.~\ref{fig:network}(a). For a high-quality image $x_h$, there is no artifact and therefore only the content $c_h$ is encoded.

Our network takes two unpaired images $x_l \in \mathbb{X}_l$ and $x_h \in \mathbb{X}_h$ as inputs. For $x_l$, we use a content encoder $E_l$ and an artifact encoder $E_a$ to encode its content and artifact components, respectively. As $x_h$ does not contain artifacts, we only use a content encoder $E_h$ to encode its content. The latent codes are written as,
$c_l = E_l(x_l), c_h = E_h(x_h), a = E_a(x_l)$.

This disentanglement allows decodings among the different combinations of the artifact and content components of $\mathbb{X}_l$ and $\mathbb{X}_h$, which enable four generators $x_{l \rightarrow l}$, $x_{l \rightarrow h}$, $x_{h \rightarrow l}$, and $x_{h \rightarrow h}$. $x_{i \rightarrow j}$ means that the output is encoded with the content of $x_i, i \in \{l, h\}$ and intended to look like a sample from $\mathbb{X}_j, j \in \{l, h\}$. We use two different generators $G_l$ and $G_h$ for each image domain. The low-quality image generator $G_l$ takes a content code $c_i, i \in \{l, h\}$ and an artifact code $a$ as inputs and outputs a low-quality image $x_{i \rightarrow l}$:
\begin{equation}
x_{l \rightarrow l} = G_l(c_l, a), x_{h \rightarrow l} = G_l(c_h, a).
\end{equation}

The high-quality image generator $G_h$ takes a content code $c_i, i \in \{l, h\}$ and a shape attention map $m_i$ as inputs and outputs a high-quality image $x_{i \rightarrow h}$:
\begin{equation}
x_{l \rightarrow h} = G_h(c_l, m_l), x_{h \rightarrow h} = G_h(c_h, m_h),
\end{equation}
where the shape attention map $m_i = G_s(c_i)$ is generated by a shape generator $G_s$. We use $m_i$ to explicitly fuse the vertebra shape information into the decoding such that $G_h$ (Fig.~\ref{fig:network}(c)) generates better the vertebra region, which is critical in clinical practice. We will also show later (Section~\ref{sec:learning}) that learning $m_i$ can be achieved using the vertebra shapes $y_h^{gt}$ from $G_h$. 

One goal of this work is to segment vertebra shapes $\mathbb{Y}$ from $\mathbb{X}_l$. Hence, we use a low-quality image segmentor $S_l$ and a high-quality image segmentor $S_h$ to map images from domain $\mathbb{X}_l$ and domain $\mathbb{X}_h$ to space $\mathbb{Y}$, respectively:
\begin{equation}
\begin{aligned}
&y_l = S_l(x_l), y_{l \rightarrow l} = S_l(x_{l \rightarrow l}), y_{h \rightarrow l} = S_l(x_{h \rightarrow l}),\\
&y_h = S_h(x_h), y_{h \rightarrow h} = S_h(x_{h \rightarrow h}), y_{l\rightarrow h} = S_h(x_{l \rightarrow h}).
\end{aligned}
\end{equation}

\subsection{Network learning and loss functions} \label{sec:learning}
To promote network learning, we design image- and shape-domain losses that leverage the adversarial costs as well as various forms of consistency between the inputs and outputs to obviate the need for the groundtruth data of $x_l$. 

\textit{Image domain losses} encourage the network to generate the four outputs $\{x_{i \rightarrow j} \mid i \in \{l, h\}, j \in \{l, h\}\}$ as intended, i.e., $x_{i \rightarrow j}$ should match the content of $x_i$ and look like a sample from $\mathbb{X}_j$. We use $L_1$ loss to regularize in-domain reconstruction and adversarial losses to encourage cross-domain translation,
\begin{equation}
\mathcal{L}_{recon} = \mathbb{E}_{\mathbb{X}_l,\mathbb{X}_h}[||x_l-x_{l \rightarrow l}||_1+||x_h-x_{h \rightarrow h}||_1], 
\end{equation}
\begin{equation}
\begin{aligned}
\mathcal{L}_{adv} &= \mathbb{E}_{\mathbb{X}_l}[\log D_l(x_l)] + \mathbb{E}_{\mathbb{X}_l,\mathbb{X}_h}[1- \log D_l(x_{h \rightarrow l})] \\ & + \mathbb{E}_{\mathbb{X}_h}[\log D_h(x_h)] + \mathbb{E}_{\mathbb{X}_l,\mathbb{X}_h}[1- \log D_h(x_{l \rightarrow h})].
\end{aligned}
\end{equation}

To ensure the artifacts generated in $x_{h \rightarrow l}$ can also be removable by our model, we apply the cycle consistency loss~\cite{zhu2017unpaired,huang2018multimodal} for $x_h \rightarrow x_{h \rightarrow l} \rightarrow x_{h \rightarrow l \rightarrow h}$:
\begin{equation}
\mathcal{L}_{cycle} = \mathbb{E}_{\mathbb{X}_l,\mathbb{X}_h}[|| G_h(E_l(x_{h \rightarrow l}), G_s(E_l(x_{h \rightarrow l})))-x_h||_1].
\end{equation}

To further impose the anatomy preciseness, we employ an \textit{artifact consistency loss}~\cite{liao2019adn}, ensuring that the same artifact is removed from $x_l$ and added to $x_{h \rightarrow l}$:
\begin{equation}
\mathcal{L}_{arti} = \mathbb{E}_{\mathbb{X}_l,\mathbb{X}_h}[||(x_l-x_{l \rightarrow h})-(x_{h \rightarrow l}-x_h)||_1].
\end{equation}

\textit{Shape domain losses} leverage the ground truth vertebra shape $y_h^{gt}$ in CT domain and shape consistencies for the learning of two segmentors $S_l$ and $S_h$. Based on Dice loss $\delta$ against $y_h^{gt}$, explicit shape constraints can be applied on the segmentation maps $\{y_h,y_{h \rightarrow l},y_{h \rightarrow h}\}$ and the decoded attention map $m_h$ of the image $x_h$:  
\begin{equation}
\mathcal{L}_{segm} =  \mathbb{E}_{\mathbb{X}_l,\mathbb{X}_h}[\delta(y_h)+\delta(y_{h \rightarrow l})+\delta(y_{h \rightarrow h})]; ~~ \mathcal{L}_{segm}^m = \mathbb{E}_{\mathbb{X}_h}[\delta(m_h)].
\end{equation}

As the anatomical information is supposed to be retained during image reconstruction and translation, we employ \textit{anatomy consistency} losses to minimize the distance of their segmentation results,
\begin{equation}
\mathcal{L}_{anat} = \mathbb{E}_{\mathbb{X}_l,\mathbb{X}_h}[||y_l-y_{l \rightarrow l}||_1+||y_h-y_{h \rightarrow h}||_1 + ||y_l-y_{l \rightarrow h}||_1+||y_h-y_{h \rightarrow l}||_1].
\end{equation}

The overall objective function is the weighted sum of all the above losses, we set the weight of $\mathcal{L}_{adv}$ to 1, and all the other weights to 5. We set the weights of losses based on the importance of each component and experimental experience.

\subsection{Anatomy-aware modality translation} \label{section:generator}
To better retain the anatomical structure in the synthetic CT image, we integrate the anatomy knowledge into the idea of SPADE \cite{park2019semantic} and form an \textbf{anatomy-aware de-normalization (AADE) layer} (see Fig.~\ref{fig:network}(b)). AADE first normalizes the input feature $F_{in}$ with a scale $\sigma$ and a shift $\mu$ using a parameter-free batch-normalization (Norm) layer, and then denormalizes it based on a shape attention map $m_i, i \in \{l, h\}$ through learn-able parameters $\gamma$ and $\beta$:
\begin{equation}
F_{out} =\frac{F_{in}-\mu(F_{in})}{\sigma(F_{in})}\times \gamma(\mathcal{R}(m_i)) + \beta(\mathcal{R}(m_i)).
\end{equation}
where $\mathcal{R}$ resamples $m_i$ to the spatial dimension of $F_{in}$, and $F_{out}$ denotes the output feature map. $\gamma$ and $\beta$ are learned from $m_i$ by three Conv layers. The first Conv layer encodes $\mathcal{R}(m_i)$ to a hidden space and then the other two Conv layers learn spatially related parameter $\gamma$ and $\beta$, respectively. 
All the Norm layers in residual, upsampling and final blocks of $G_h$ are replaced by the AADE layer. Our model benefits from the new structure in two aspects. First, the learned shape representation guides the synthesis, which prevents washing away the anatomical information. Second, the soft mask allows the gradients to be back-propagated through disentanglement learning, which encourages the encoding of content code to be more accurate. 

\begin{figure*}[t]
	\begin{center}
		\includegraphics[width=0.9\linewidth]{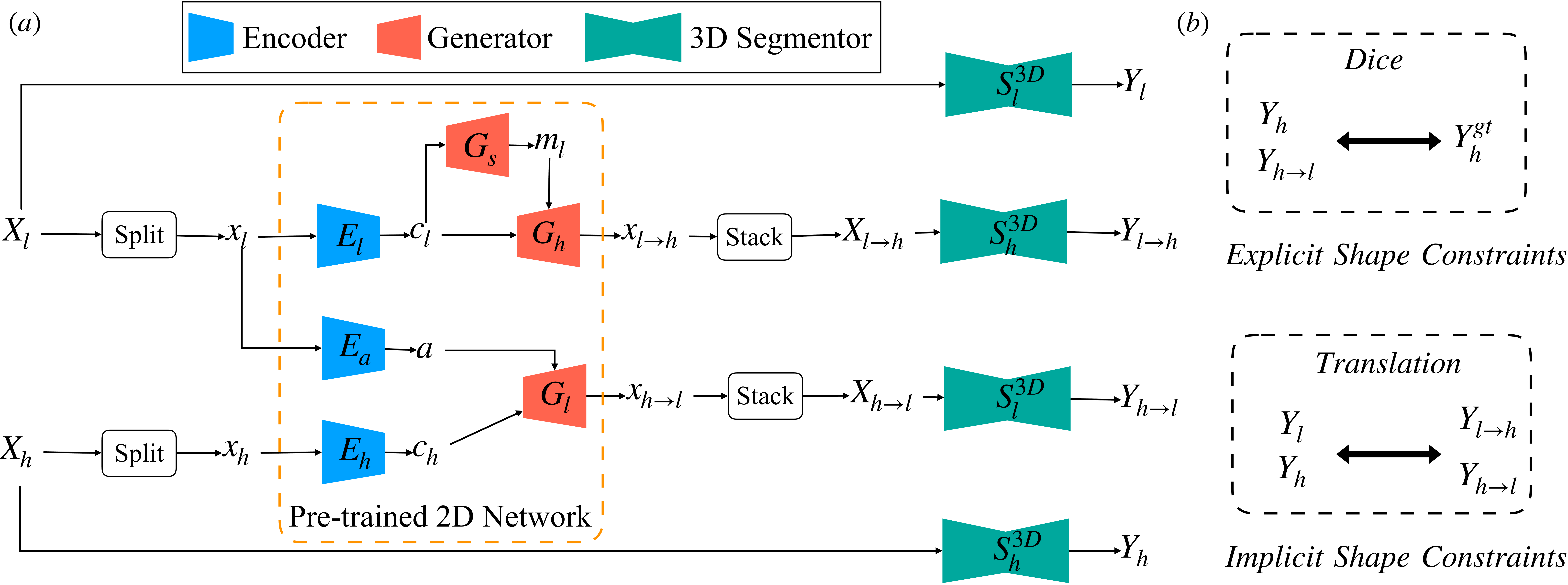}
	\end{center}
	\caption{(a) Anatomy-aware 3D segmentation pipeline. (b) Shape constraints.}
	\label{fig:network_3d}
\end{figure*}

\subsection{3D segmentation}
As the above network is designed for 2D images, it is difficult to keep inter-slice spatial continuity. As in Fig.~\ref{fig:network_3d}(a), we design a 3D segmentation network. A 3D CBCT volume $X_l$ and a CT volume $X_h$ are first split to 2D images $x_l$ and $x_h$, then the pre-trained A$^3$DSegNet network serves as an online translator and generates synthetic images $x_{l \rightarrow h}$ and  $x_{h \rightarrow l}$. We recombine $x_{l \rightarrow h}$ and  $x_{h \rightarrow l}$ to $X_{l \rightarrow h}$ and  $X_{h \rightarrow l}$ by stacking them along the slice dimension. Finally, we use a low-quality 3D segmentor $S_l^{3D}$ and a high-quality 3D segmentor $S_h^{3D}$ to map a volume from image domain to shape domain:
\begin{equation} \begin{aligned}
Y_l = S_l^{3D}(X_l), Y_{h \rightarrow l} = S_l^{3D}(X_{h \rightarrow l}), Y_h = S_h^{3D}(X_h), Y_{l\rightarrow h} = S_h^{3D}(X_{l \rightarrow h}). 
\end{aligned}
\end{equation}
Since the groundtruth segmentation of CBCT image is not available, we apply explicit shape constraints on CT segmentation and implicit shape constraints between raw and translated segmentation results (Fig.~\ref{fig:network_3d} (b)):
\begin{equation}
\mathcal{L}^{3D}_{segm} = \mathbb{E}_{\mathbb{X}_h}[\delta(Y_h)+\delta(Y_{h \rightarrow l})],
\end{equation}
\begin{equation}
\mathcal{L}^{3D}_{anat} = \mathbb{E}_{\mathbb{X}_l,\mathbb{X}_h}[||Y_l-Y_{l \rightarrow h}||_1+||Y_h-Y_{h \rightarrow l}||_1].
\end{equation}

\section{Experiments}

\subsection{Dataset and experiment setup}

\textbf{CBCT data.}
The CBCT data were collected by a Siemens Arcadis Orbic 3D system during spinal intervention. The dataset contains 109 CBCT scans and all of the scans cover two or three lumbar vertebrae. The size of CBCT volumes is 256 $\times$ 256 $\times$ 256. The isotropic voxel size is 0.5 mm. Due to the severe cone-beam geometry distortion at the two ends of the sagittal axis, we only keep 196 slices in the middle for each volume. We use 97 volumes for training and 12 volumes for testing, resulting in 19,012 slices in the training set and 2,352 slices in the testing set. To evaluate the segmentation performance, the vertebra masks for the testing set only were manually labeled by an expert.

\noindent\textbf{CT data.}
We include four public datasets as high-quality CT images:  Dataset 13 and Dataset 15 of SpineWeb~\cite{glocker2013vertebrae,ibragimov2017segmentation,yao2016multi}, VerSe 19 \cite{sekuboyina2020verse}, UL dataset \cite{ibragimov2015interpolation}. In total, we include 125 CT scans with corresponding segmentation masks for the vertebrae. The in-plane resolution is [0.31,1.0]~mm and the slice thickness is [0.7,3.0]~mm. We only include the CT slices of the lumbar vertebrae in the experiment. To match the resolution and spatial dimension of CBCT image, all the CT images are resampled to a spacing of 0.5$\times$0.5$\times$1~mm$^3$.  We use 105 scans for training and 20 scans for testing with the same training/testing ratio for every dataset, resulting in 14,737 images for training and 2,352 images for testing. 

\noindent\textbf{Implementation details.} We implement our model using the PyTorch framework. For 2D disentanglement network, we train it for 15 epochs using the Adam optimizer with a learning rate of $1\times10^{-4}$ and a batch size of 1. For 3D segmentation network, the size of the input patch is 96$\times$256$\times$256 and we downsample the patch to a spacing of 1$\times$1$\times$1~mm$^3$ after translation to save GPU memory. We train the 3D segmentation network for 78,386 iterations. 

\noindent\textbf{Metrics.} We evaluate the performance of vertebra segmentation using the Dice score and the average symmetric surface distance (ASD). We obtain the segmentation mask from shape prediction by applying a threshold of 0.5. A higher Dice and a lower ASD mean a better segmentation performance.


\subsection{Ablation study}
In this section, we investigate the effectiveness of different modules and objectives of the proposed architecture. Here we focus on the A$^3$DSegNet network using $E_l$, $E_h$, $E_a$, $G_l$, $G_h$, $G_s$ and learning with image domain losses ($\mathcal{L}_{adv}$, $ \mathcal{L}_{recon}$, $\mathcal{L}_{cycle}$, and $ \mathcal{L}_{arti} $) and shape domain losses ($ \mathcal{L}_{segm}$, $ \mathcal{L}^{m}_{segm}$, and $ \mathcal{L}_{anat}$). The configurations of different models are as follows: 

\begin{itemize}
    \item M$_1$: $\{E_l, E_h, E_a, G_l, G_h\}$ + $\{S_l, S_h\}$, using $\mathcal{L}_{recon}$, $\mathcal{L}_{adv}$, $\mathcal{L}_{cycle}$, and $ \mathcal{L}_{segm}$;
    \item M$_2$: M$_1$ but using $ \mathcal{L}_{anat}$ as an additional loss;
    \item M$_3$: M$_2$ + $\{G_s\}$, using AADE and $ \mathcal{L}^{m}_{segm}$ as an additional loss;
    \item M$_4$ (full): M$_3$ without using $ \mathcal{L}_{arti}$.
\end{itemize}

\begin{figure*}[t]
	
	\centering
	\begin{minipage}[t]{1\textwidth}
	\begin{minipage}[t]{0.13\textwidth}
	\centering
	$x_l$/$x_h$/$y^{gt}_{l}$
	\end{minipage}
	\begin{minipage}[t]{0.13\textwidth}
	\centering
	$x_{l \rightarrow h}$
	\end{minipage}
	\begin{minipage}[t]{0.13\textwidth}
	\centering
	$x_{h \rightarrow l}$
	\end{minipage}
	\begin{minipage}[t]{0.13\textwidth}
	\centering
	$y_l$
	\end{minipage}
	\begin{minipage}[t]{0.13\textwidth}
	\centering
	$y_{l \rightarrow l}$
	\end{minipage}
	\begin{minipage}[t]{0.13\textwidth}
	\centering
	$y_{l \rightarrow h}$
	\end{minipage}
	\begin{minipage}[t]{0.13\textwidth}
	\centering
	$m_l$
	\end{minipage}
	\begin{minipage}[t]{0.03\textwidth}
	\hspace*{0.2cm}
	\end{minipage}
	\end{minipage}
	
	\begin{minipage}[b]{1\textwidth}
	\begin{minipage}[t]{0.13\textwidth}
		\centering 
		\includegraphics[width=1\textwidth]{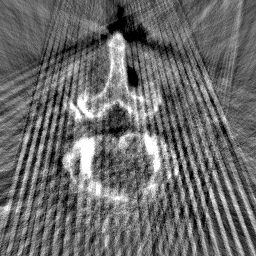}
	\end{minipage}
	\begin{minipage}[t]{0.13\textwidth}
		\centering
		\includegraphics[width=1\textwidth]{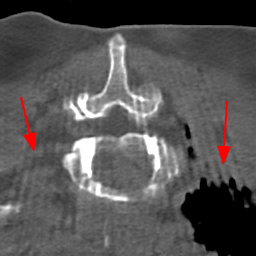}
	\end{minipage}
	\begin{minipage}[t]{0.13\textwidth}
		\centering
		\includegraphics[width=1\textwidth]{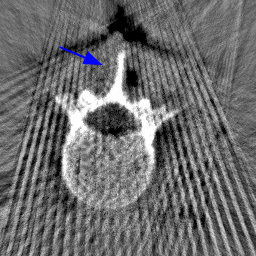}
	\end{minipage}
	\begin{minipage}[t]{0.13\textwidth}
		\centering
		\includegraphics[width=1\textwidth]{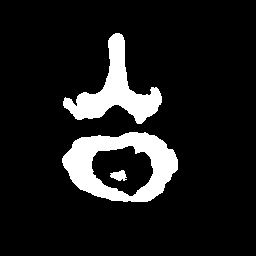}
	\end{minipage}
	\begin{minipage}[t]{0.13\textwidth}
		\centering
		\includegraphics[width=1\textwidth]{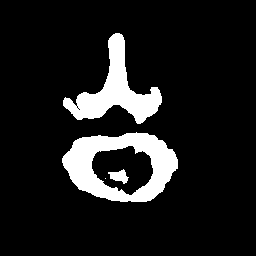}
	\end{minipage}
	\begin{minipage}[t]{0.13\textwidth}
		\centering
		\includegraphics[width=1\textwidth]{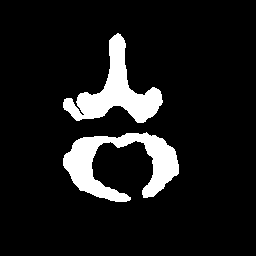}
	\end{minipage}
	\begin{minipage}[t]{0.13\textwidth}
		\centering
		\includegraphics[width=1\textwidth]{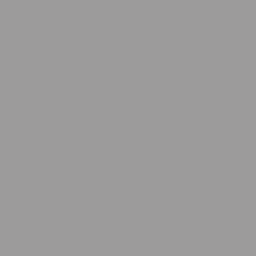}
	\end{minipage}
	\begin{minipage}[b]{0.03\textwidth}
		M$_1$
		\vspace*{0.2cm}
	\end{minipage}
	\end{minipage}
	
	\begin{minipage}[t]{1\textwidth}
    \begin{minipage}[t]{0.13\textwidth}
		\includegraphics[width=1\textwidth]{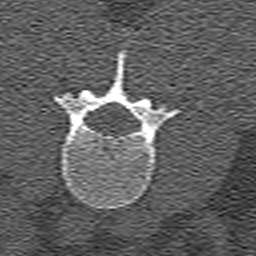}
	\end{minipage}
	\begin{minipage}[t]{0.13\textwidth}
		\includegraphics[width=1\textwidth]{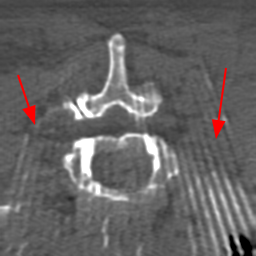}
	\end{minipage}
	\begin{minipage}[t]{0.13\textwidth}
		\includegraphics[width=1\textwidth]{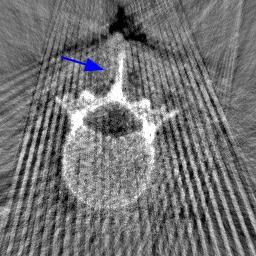}
	\end{minipage}
	\begin{minipage}[t]{0.13\textwidth}
		\includegraphics[width=1\textwidth]{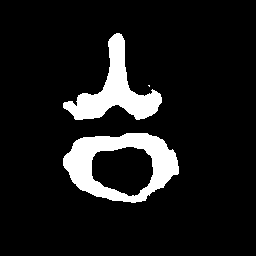}
	\end{minipage}
	\begin{minipage}[t]{0.13\textwidth}
		\includegraphics[width=1\textwidth]{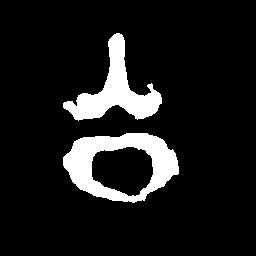}
	\end{minipage}
	\begin{minipage}[t]{0.13\textwidth}
		\includegraphics[width=1\textwidth]{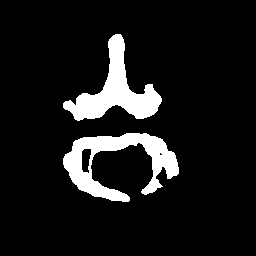}
	\end{minipage}
	\begin{minipage}[t]{0.13\textwidth}
		\includegraphics[width=1\textwidth]{figures/ablation/white.png}
	\end{minipage}
	\begin{minipage}[b]{0.03\textwidth}
		M$_2$
		\vspace*{0.2cm}
	\end{minipage}
	\end{minipage}
	
	\begin{minipage}[t]{1\textwidth}
	\begin{minipage}[t]{0.13\textwidth}
		\includegraphics[width=1\textwidth]{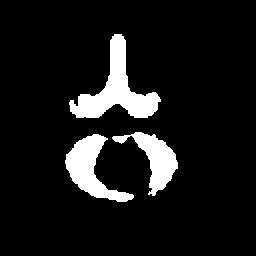}
	\end{minipage}
	\begin{minipage}[t]{0.13\textwidth}
		\includegraphics[width=1\textwidth]{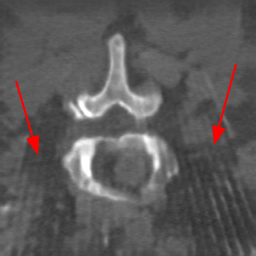}
	\end{minipage}
	\begin{minipage}[t]{0.13\textwidth}
		\includegraphics[width=1\textwidth]{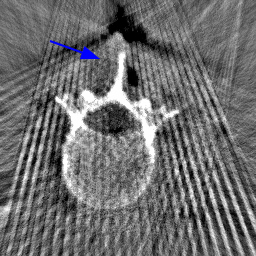}
	\end{minipage}
	\begin{minipage}[t]{0.13\textwidth}
		\includegraphics[width=1\textwidth]{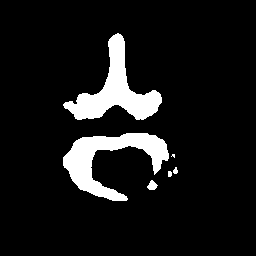}
	\end{minipage}
	\begin{minipage}[t]{0.13\textwidth}
		\includegraphics[width=1\textwidth]{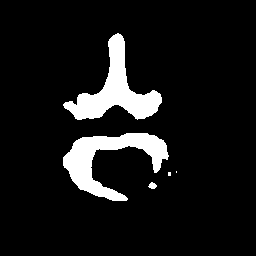}
	\end{minipage}
	\begin{minipage}[t]{0.13\textwidth}
		\includegraphics[width=1\textwidth]{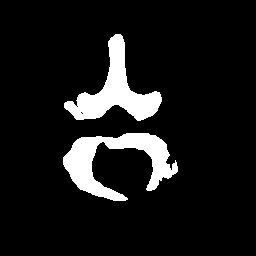}
	\end{minipage}
	\begin{minipage}[t]{0.13\textwidth}
		\includegraphics[width=1\textwidth]{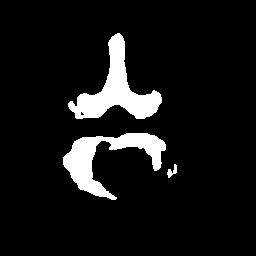}
	\end{minipage}
	\begin{minipage}[b]{0.03\textwidth}
		M$_3$
		\vspace*{0.2cm}
	\end{minipage}
	\end{minipage}

	\begin{minipage}[t]{1\textwidth}
	\begin{minipage}[t]{0.13\textwidth}
		\includegraphics[width=1\textwidth]{figures/ablation/white.png}
	\end{minipage}
	\begin{minipage}[t]{0.13\textwidth}
		\includegraphics[width=1\textwidth]{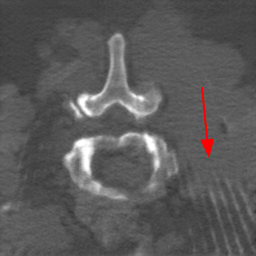}
	\end{minipage}
	\begin{minipage}[t]{0.13\textwidth}
		\includegraphics[width=1\textwidth]{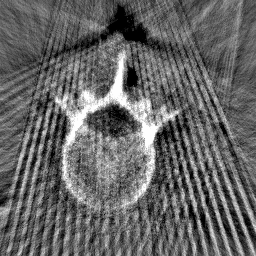}
	\end{minipage}
	\begin{minipage}[t]{0.13\textwidth}
		\includegraphics[width=1\textwidth]{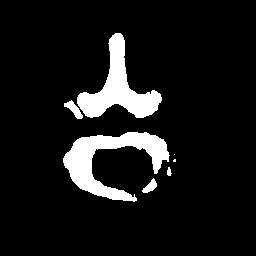}
	\end{minipage}
	\begin{minipage}[t]{0.13\textwidth}
		\includegraphics[width=1\textwidth]{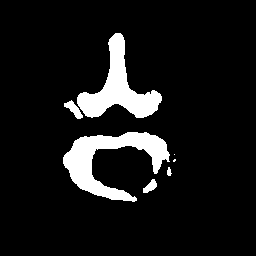}
	\end{minipage}
	\begin{minipage}[t]{0.13\textwidth}
		\includegraphics[width=1\textwidth]{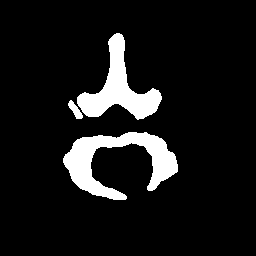}
	\end{minipage}
	\begin{minipage}[t]{0.13\textwidth}
		\includegraphics[width=1\textwidth]{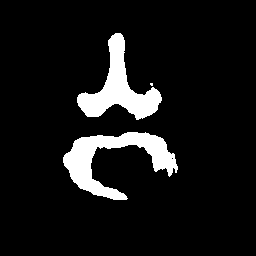}
	\end{minipage}
	\begin{minipage}[b]{0.03\textwidth}
		M$_4$
		\vspace*{0.2cm}
	\end{minipage}
	\end{minipage}

	\caption{Visual comparison for synthetic image and segmentation of different models. M$_4$ (full) produces the least amount of artifact and the most complete segmentation.}
	\label{fig:ablation}
\end{figure*}

\begin{table*} [t]
	\caption{Quantitative evaluation segmentation performance for different models.}
	\begin{center}
		\resizebox{\linewidth}{!}{
			\begin{tabular}{l|rrrrrrrr}
				\toprule[1pt]
				\footnotesize{Dice/ASD(mm)}&$m_l$&$y_l$&$y_{l \rightarrow l}$&$y_{l \rightarrow h}$&$m_h$&$y_h$&$y_{h \rightarrow h}$&$y_{h \rightarrow l}$\\
				\hline
				\hline
				M$_1$&n.a./n.a.& .802/2.16& .801/2.17& .806/2.12&n.a./n.a.& .925/0.89& .918/0.98& .911/1.11\\
				M$_2$&n.a./n.a.& .806/1.94& .805/1.96& .806/2.04&n.a./n.a.& .930/0.80& .928/0.83& .923/\textbf{0.89}\\
				M$_3$&.774/2.58&.828/1.68&.828/1.69&.836/1.64&\textbf{.918}/\textbf{1.14}&.930/0.79&.929/0.80&.923/1.00\\
				M$_4$&\textbf{.829}/\textbf{2.01}&\textbf{.843}/\textbf{1.57}&\textbf{.843}/\textbf{1.56}&\textbf{.847}/\textbf{1.54}&.915/1.18&\textbf{.932}/\textbf{0.79}&\textbf{.932}/\textbf{0.79}&\textbf{.925}/0.94\\
				\toprule[1pt]
			\end{tabular}
		}
	\end{center}
	\label{table:ablation}\vspace{-0.3in}
\end{table*}

\noindent\textbf{Disentanglement and explicit shape constraints.} As shown in Fig.~\ref{fig:ablation}, we can see streak metal artifacts nearly everywhere in $x_l$. M$_1$ can roughly disentangle artifacts and anatomical information but strong vertical artifacts and strange air area appear in $x_{l \rightarrow h}$ (see red arrows of M$_1$ in Fig.~\ref{fig:ablation}). For the anatomical structure, M$_1$ learns to segment vertebrae with fully supervised $S_h$ and $S_l$ applied on various CT images, but fails to suppress the false bony structure in $y_l$ and $y_{l \rightarrow l}$ as $S_l$ may misclassify some metal artifacts as bone.

\noindent\textbf{Implicit shape constraints.} With $ \mathcal{L}_{anat}$, all segmentations are improved with higher Dices and smaller ASDs, see Table~\ref{table:ablation}. As shown in Fig.~\ref{fig:ablation}, $y_{l}$, $y_{l \rightarrow l}$ and $y_{l \rightarrow h}$ become similar but the high density bone is not correctly segmented in $y_{l \rightarrow h}$ as it maybe treated as metal artifacts. Comparing $x_{l \rightarrow h}$ between M$_2$ and M$_1$, the abnormal air region disappears but metal artifact reduction performance is still not satisfactory.

\noindent\textbf{Anatomy-aware generation.} With AADE layer in M$_3$, $y_l$, $y_{l \rightarrow l}$ and $y_{l \rightarrow h}$ are substantially improved as shown in Table~\ref{table:ablation}. Note, $m_l$ is used as attention map, so we do not expect it to be identical to $y_l^{gt}$.  In $x_{l \rightarrow h}$, metal artifacts are further suppressed comparing with M$_2$. Thus, AADE is critical to our anatomy-aware artifact disentanglement framework. With the special structure, $G_s$ can be punished in the image translation and reconstruction processes and the other encoders and generators receive more guidance. However, as shown by blue arrows in Fig.~\ref{fig:ablation}, we observe a shadow of vertebra edge of $x_l$ appears in $y_{l \rightarrow h}$ of M$_3$ and the vertebra boundaries get smoothed out in $y_{h \rightarrow l}$. It may be because sharp edges are encoded as metal artifacts and forced to be added to $y_{h \rightarrow l}$ by artifact consistency loss $ \mathcal{L}_{arti} $.

\noindent\textbf{Removal of $ \mathcal{L}_{arti} $.}  To mitigate vertebrae shadows, we remove $ \mathcal{L}_{arti} $. The segmentation performance of most images in M$_4$ gets improved because of better synthetic images. Overall, $y_{l \rightarrow h}$ in M$_4$ yields the best segmentation performance for CBCT images with an average Dice of 0.847 and an average ASD of 1.54 mm. For the synthetic images, M$_4$ generates $x_{l \rightarrow h}$ with the best quality and least metal artifacts among all the models. M$_4$ also outputs $x_{h \rightarrow l}$ without vertebra shadows. The results indicate our shape-aware network could preserve anatomical details and transfer the metal artifacts precisely without $\mathcal{L}_{arti}$.



\subsection{Comparison with state-of-the-art}
We compare our model with competing methods to show the benefits of joint learning of segmentation, artifact reduction, and modality translation tasks.

\begin{figure*}[t]
	\begin{minipage}[t]{0.5\textwidth}
		\centering
		\begin{minipage}[t]{0.18\textwidth}
			\centering
			\includegraphics[width=1\textwidth]{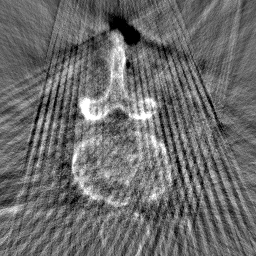}
			\includegraphics[width=1\textwidth]{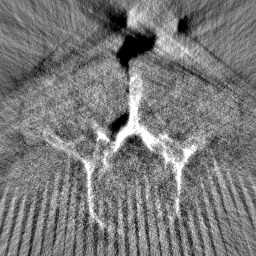}    
			\includegraphics[width=1\textwidth]{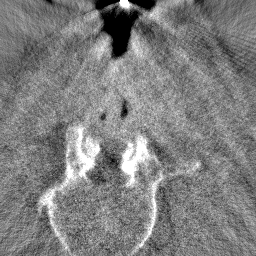}
			\includegraphics[width=1\textwidth]{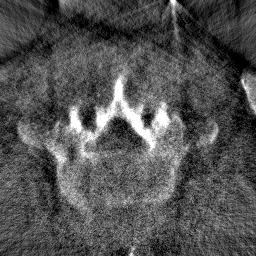}
			
			$x_l$
		\end{minipage}
		\begin{minipage}[t]{0.18\textwidth}
			\centering
			\includegraphics[width=1\textwidth]{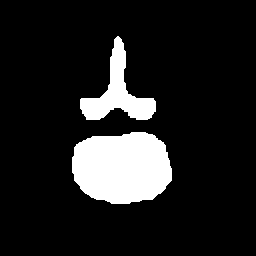}
			\includegraphics[width=1\textwidth]{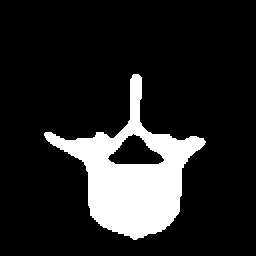}
			\includegraphics[width=1\textwidth]{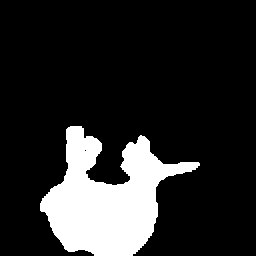}
			\includegraphics[width=1\textwidth]{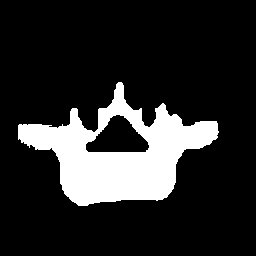}
			
			$y_l^{gt}$
		\end{minipage}
		\begin{minipage}[t]{0.18\textwidth}
			\centering
			\includegraphics[width=1\textwidth]{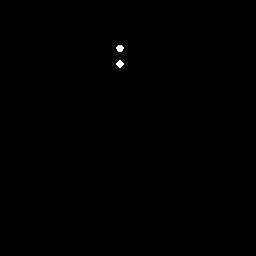}
			\includegraphics[width=1\textwidth]{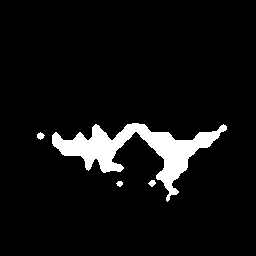}
			\includegraphics[width=1\textwidth]{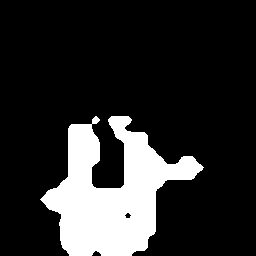}
			\includegraphics[width=1\textwidth]{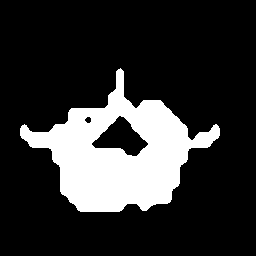}			
			\scriptsize{AdaptSeg-Net}
		\end{minipage}
		\begin{minipage}[t]{0.18\textwidth}
			\centering
			\includegraphics[width=1\textwidth]{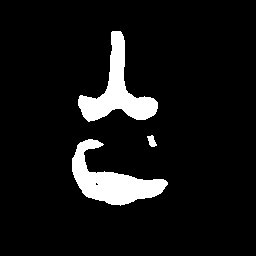}
			\includegraphics[width=1\textwidth]{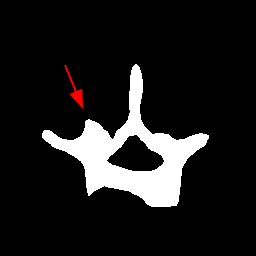}    
			\includegraphics[width=1\textwidth]{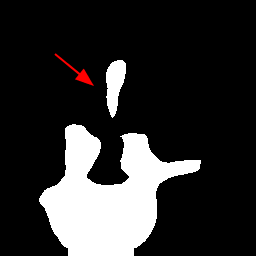}
			\includegraphics[width=1\textwidth]{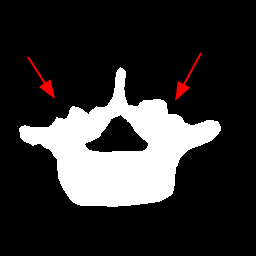}	
			\scriptsize{SIFA}
		\end{minipage}
		\begin{minipage}[t]{0.18\textwidth}
			\centering
			\includegraphics[width=1\textwidth]{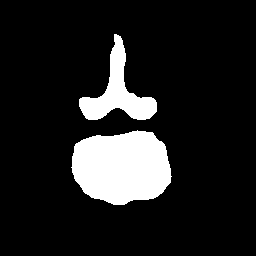}
			\includegraphics[width=1\textwidth]{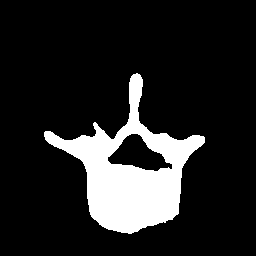}
			\includegraphics[width=1\textwidth]{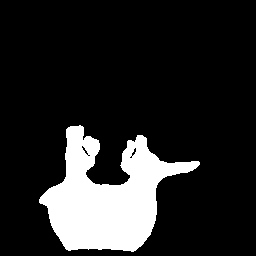}
			\includegraphics[width=1\textwidth]{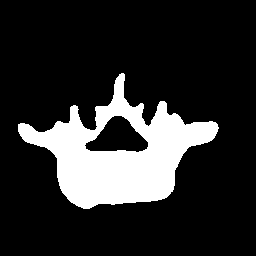}
			Ours
		\end{minipage}
		\caption{Visualization for $x_l$ and segmentation results of competing methods on two cases.}
		\label{fig:eval_seg}
	\end{minipage}
	\begin{minipage}[t]{0.5\textwidth}
		\centering
		\begin{minipage}[t]{0.18\textwidth}
			\centering
			\includegraphics[width=1\textwidth]{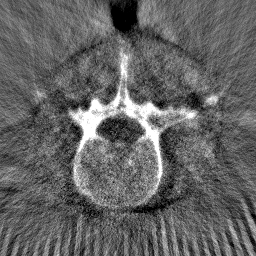}
			\includegraphics[width=1\textwidth]{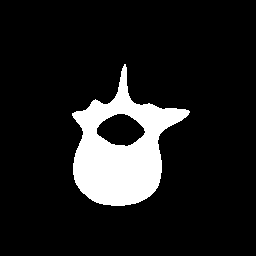}
			\includegraphics[width=1\textwidth]{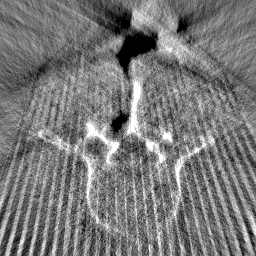}
			\includegraphics[width=1\textwidth]{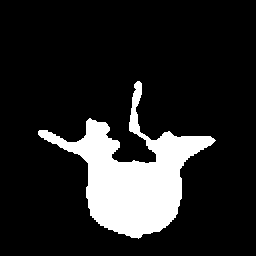}
			$x_l$
		\end{minipage}
		\begin{minipage}[t]{0.18\textwidth}
			\centering
			\includegraphics[width=1\textwidth]{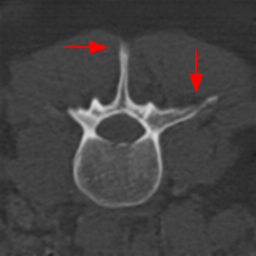}
			\includegraphics[width=1\textwidth]{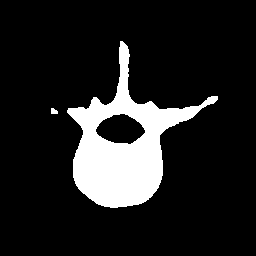}
			\includegraphics[width=1\textwidth]{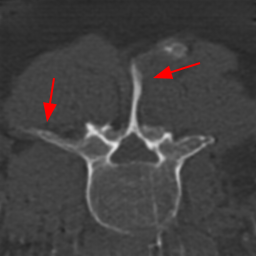}
			\includegraphics[width=1\textwidth]{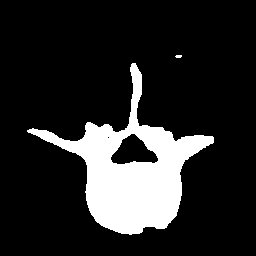}
			\scriptsize{Cycle-GAN}
		\end{minipage}
		\begin{minipage}[t]{0.18\textwidth}
			\centering
			\includegraphics[width=1\textwidth]{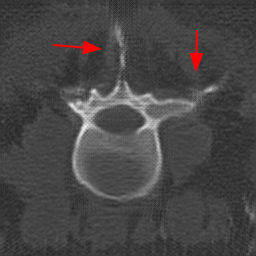}
			\includegraphics[width=1\textwidth]{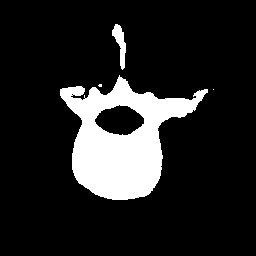}
			\includegraphics[width=1\textwidth]{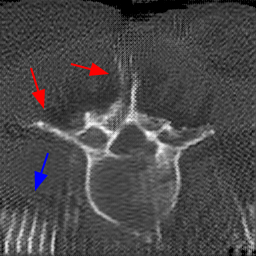}
			\includegraphics[width=1\textwidth]{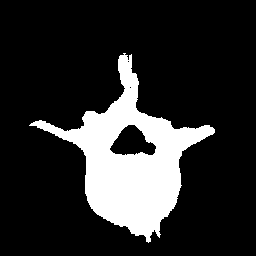}
			DRIT 
		\end{minipage}
		\begin{minipage}[t]{0.18\textwidth}
			\centering
			\includegraphics[width=1\textwidth]{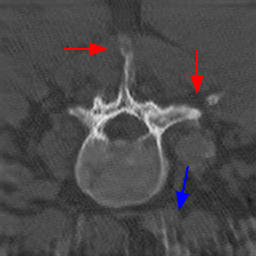}
			\includegraphics[width=1\textwidth]{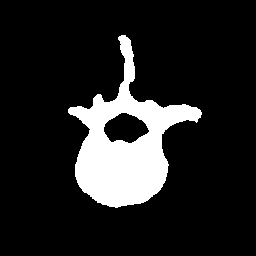}
			\includegraphics[width=1\textwidth]{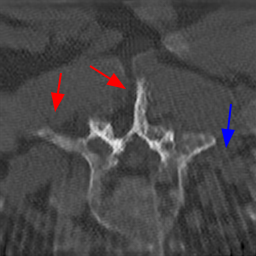}
			\includegraphics[width=1\textwidth]{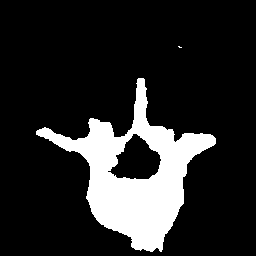}
			ADN 
		\end{minipage}
		\begin{minipage}[t]{0.18\textwidth}
			\centering
			\includegraphics[width=1\textwidth]{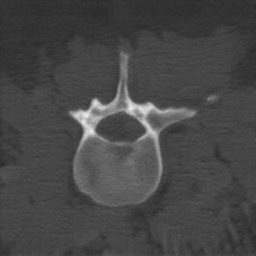}
			\includegraphics[width=1\textwidth]{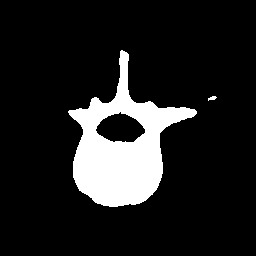}
			\includegraphics[width=1\textwidth]{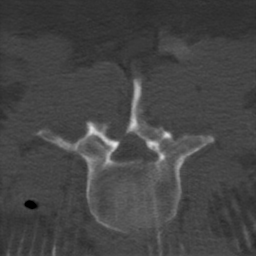}
			\includegraphics[width=1\textwidth]{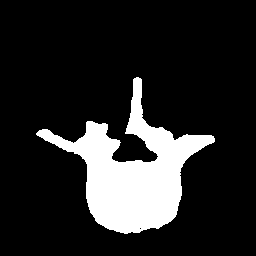}
			Ours
		\end{minipage}
		\caption{Visualization for modality translation of two cases. Odd rows are $x_{l \rightarrow h}$, and even rows are segmentations outputed by a pre-trained UNet on CT images. }
		\label{fig:eval_syn}
	\end{minipage}
\end{figure*}

\noindent\textbf{2D segmentation.}
Our model is compared with two methods based on domain adaptation: AdaptSegNet \cite{Tsai_adaptseg_2018} and SIFA \cite{chen2020unsupervised}. AdaptSegNet and SIFA are trained with the officially released codes. AdaptSegNet is trained with Dice loss as we only have one class here. The results are summarized in Fig.~\ref{fig:eval_seg} and Table~\ref{table:eval_seg}(a). AdaptSegNet invokes DeeplabV2 as the segmentor and cannot capture the vertebra especially when metal artifacts exist. SIFA outputs plausible predictions but the performance is heavily affected by the metal artifacts. Also, the segmentations predicted by SIFA can not capture vertebrae precisely and show false positive bones and enlarged masks (see red arrows in Fig.~\ref{fig:eval_seg}). With joint learning, our model achieves the best segmentation performance with an average Dice of 0.847 and an average ASD of 1.54mm.


\noindent\textbf{Modality translation and artifact reduction.}
Here we compare our model with other methods: CycleGAN \cite{zhu2017unpaired}, DRIT \cite{lee2018diverse}, ADN \cite{liao2019adn}. All the models are trained with our data using their officially released codes. Further, we train a UNet segmentation network using annotated CT data and apply it to synthesized CT images as an anatomy-invariant segmentation evaluator.
As shown in Table~\ref{table:eval_seg}(b), our model achieves the best performance with a much larger average Dice compared with other methods. Fig.~\ref{fig:eval_syn} shows the synthetic images and segmentation results. CycleGAN and DRIT tend to output plausible and realistic CT images but are not able to preserve the anatomical information precisely. As shown by the red arrows in Fig.~\ref{fig:eval_syn}, the bony structures appear distorted and noisy. ADN can retain most of the anatomical information but not for the bone pixels with high intensity, which might be classified into metal artifacts. With anatomical knowledge learned from the CT domain, our model outputs high-quality synthetic CT images while keeping anatomical consistency.

For artifact reduction, ADN and DRIT \cite{lee2018diverse} could not successfully recover the clean images and streak artifacts remain in the synthetic image (see blue arrows in Fig.~\ref{fig:eval_syn}). CycleGAN \cite{zhu2017unpaired} could output clean images but the distorted bones make them less valuable. Our model can suppress all the artifacts and keep the bone edges sharp, which outperforms all the other methods.

\begin{table*} [t]
\caption{Quantitative comparison of different competing methods for (a) segmentation, (b) modality translation.}
			\setlength{\abovecaptionskip}{0cm}
			\setlength{\belowcaptionskip}{0.15cm}
\begin{minipage}[t]{0.5\textwidth}
\begin{center}
	\resizebox{\linewidth}{!}{
		\begin{tabular}{l|r}
			\toprule[1pt]
			(a) Dice/ASD(mm)&Segmentation perf.\\
			\hline
			\hline
			AdaptSeg \cite{Tsai_adaptseg_2018}& .508/5.06\\
			SIFA \cite{chen2020unsupervised}& .825/1.93\\
			A$^3$DSegNet (ours)& \textbf{.847}/\textbf{1.54}\\
			\toprule[1pt]
		\end{tabular}
	}
\end{center}
\end{minipage}
\begin{minipage}[t]{0.43\textwidth}
	\begin{center}
		\resizebox{\linewidth}{!}{
		\begin{tabular}{l|r}
	\toprule[1pt]
(b) Dice/ASD(mm)& Segmentation perf. \\
\hline
			\hline
			CycleGAN \cite{zhu2017unpaired}&.828/2.02\\
			DRIT \cite{lee2018diverse}&.659/3.79\\
			ADN \cite{liao2019adn}&.739/2.89\\
			A$^3$DSegNet (ours) &\textbf{.846}/\textbf{1.72}\\
			\toprule[1pt]
		\end{tabular}
	}
	\end{center}
\end{minipage}
\label{table:eval_seg} \vspace{-0.1in}
\end{table*}

\begin{figure*}[t]
	\small
		\centering
		\begin{minipage}[t]{0.25\textwidth}
			\centering
			\includegraphics[width=1\textwidth]{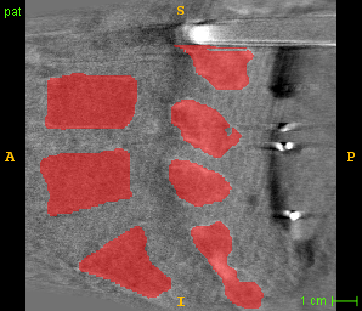}
			\includegraphics[width=1\textwidth]{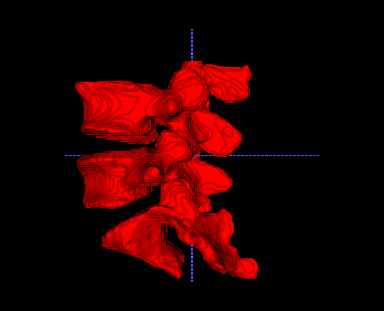}
			$X_l^{gt}$
		\end{minipage}
		\begin{minipage}[t]{0.25\textwidth}
			\centering
			\includegraphics[width=1\textwidth]{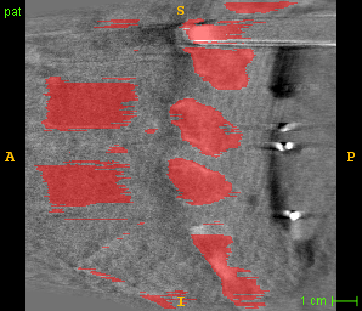}
			\includegraphics[width=1\textwidth]{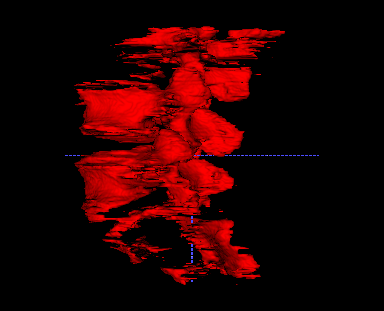}
			2D Net
		\end{minipage}
		\begin{minipage}[t]{0.25\textwidth}
			\centering
			\includegraphics[width=1\textwidth]{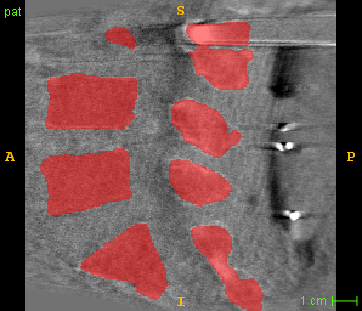}
			\includegraphics[width=1\textwidth]{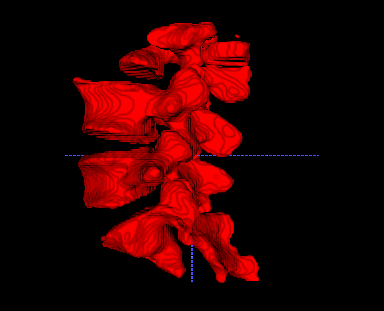}
			3D Net
		\end{minipage}
	\label{fig:3d_seg}
    \caption{Segmentation performance for 2D and 3D Nets. (Row 1): sagittal view and (2): rendering of 3D segmentation mask. The inter-slice discontinuity is fixed using 3D Net.}
\end{figure*}

To further improve the segmentation performance, we introduce the 3D segmentation network as in Fig.~\ref{fig:network_3d}. 
Our 3D model increases the Dice score from 0.819 to 0.926 and reduces the average ASD by \textbf{44\%} (from 1.47mm to 0.82mm). As shown in Fig.~\ref{fig:3d_seg}, the typical inter-slice discontinuous problem happening in 2D segmentation is fixed with our 3D model, which also runs \textbf{over 100 times} faster than slice-by slice 2D segmentation as measured on a PC with an Intel Xeon E5-2678 and a Nvidia GeForce GTX 2080 Ti.

\section{Conclusions}
To learn a vertebra segmentation model for low-quality, artifact-laden CBCT images from unpaired high-quality, artifact-free CT images with annotations, it is a must to bridge the domain and artifact gaps. To this, we present for the first time a unified framework to address three heterogeneous tasks of unpaired modality translation, vertebra segmentation, and artifact reduction. The proposed A$^3$DSegNet jointly learns content/artifact encoders, generators, and segmentors, together with an anatomy-aware de-normalization layer, through the utilization of vertebra appearance and shape knowledge across domains. Extensive results on a large number of CBCT/CT images demonstrate the effectiveness of our A$^3$DSegNet, outperforming various competing methods. In the future, we plan to conduct a clinical evaluation at multiple spinal surgery sites.
%
%
%
\bibliographystyle{splncs04}
\bibliography{egbib}

\end{document}